\documentclass[sigconf,natbib=false]{acmart}

\usepackage{amsmath}
\usepackage{caption}
\usepackage{cite} 
\usepackage{xspace}

\def\pname{{\textsc{GeoGuard}}\xspace}
\newcommand{\heading}[1]{{\vspace{2pt}\noindent\bf{#1}}} 

\setcopyright{none} 
\acmConference[Submission to AsiaCCS APKC 2026]{13th ACM ASIA Public-Key Cryptography Workshop 2026}{10 February}{TBD} 
\acmYear{2026}

\begin{document}

\title{\pname: UWB Timing-Encoded Key Reconstruction for Location-Dependent, Geographically Bounded Decryption}

\author{Kunal Mukherjee}
\email{km411@evansville.edu}
\affiliation{%
  \institution{ University of Evansville}
  \city{Evansville}
  \state{Indiana}
  \country{USA}
}





\begin{abstract} 

Digital content distribution and propitiatory research driven industries face persistent risks from intellectual property theft and unauthorized redistribution. Conventional encryption schemes such as AES, TDES, ECC, and ElGamal provide strong cryptographic guarantees, but they remain fundamentally agnostic to \emph{where} decryption takes place. In practice, this means that once a decryption key is leaked or intercepted, any adversary can misuse the key to decrypt the protected content from any location. 

This paper presents, \pname, a location-dependent cryptosystem in which the decryption key is not transmitted as data but is implicitly encoded in the precise time-of-flight differences of ultra-wideband (UWB) data transmission packets. The system leverages precise timing hardware and a custom Timing-encoded Cryptographic Keying (TiCK) protocol to map a 32-byte SHA-256 AES key onto scheduled transmission timestamps. Only user located within an approved spatial location can observe the correct packet timing that aligns with the intended packet-reception timing pattern, enabling them to reconstruct the key. Eavesdroppers outside the authorized region observe an incorrect timing pattern, which yields incorrect keys.

\pname is designed to encrypt and transmit data, but decryption is only possible when the user is within the authorized area. Our evaluation demonstrates that the system (i) removes the need to share decryption passwords electronically or physically, (ii) ensures the decryption key cannot be recovered by the eavesdropper, and (iii) provides a non-trivial spatial tolerance for legitimate users.

\end{abstract}

\maketitle

\section{Introduction}

Cryptographic primitives such as AES, DES, ECC, and ElGamal have made it possible to protect data from digital piracy and intellectual property theft~\cite{terra2016digitalpiracy, florunda2023infotheft}. These schemes are widely deployed across entertainment, cloud, financial, and research-driven industries~\cite{nist_sp80038a, ansix9_52_2016, pcidss_v4, holz2011surveyssl, koblitz1987ecccipher, elgamal1985publickey, menezes1996handbook,Mukherjee2023ProvNinja, Mukherjee2024ProvIoT,  Mukherjee2025ProvSEEK}. However, while modern ciphers offer strong guarantees on confidentiality and integrity, they are fundamentally indifferent to the \emph{geographic location} in which decryption occurs. Once an adversary obtains a valid decryption key, by intercepting it during transit or stealing it from an endpoint, there is nothing in the cryptosystem itself that prevents them from decrypting the protected content anywhere on earth unless location-based defenses are used. These industries~\cite{Mukherjee2025ProvDP, Mukherjee2025ProvExplainer} need new security measures to protect their intellectual property from corporate or international espionage.

This limitation is particularly problematic for domains where distribution is unavoidable but control is critical. Majority of intellectual property theft occurs during data transfer~\cite{miller2024techtransferrisks}. For example, the entertainment industry routinely ships high-value content (e.g., pre-release films) to theaters or partners under contractual agreements, encrypting the media in transit and then sharing keys later. The standard pattern is to encrypt content with an industry-standard cipher such as AES-256 and deliver the decryption key via a separate channel, either electronically or physically~\cite{nist_sp80057pt1r5}. This key-distribution step introduces a structural weakness: if the key is intercepted, copied, or misdirected, any unauthorized party can decrypt the content from any location.

Therefore, a viable solution is that the user should be able to decrypt a file only if he or she is at an acceptable geographic location pre-authorized by the sender. Therefore, this system will transfer the encrypted key so that only the user at the authorized location can decrypt it, without having prior knowledge of the sender or the password. Therefore, the need for sharing the password electronically or physically is omitted.

\begin{figure}[t]
  \centering
  \includegraphics[width=0.9\linewidth]{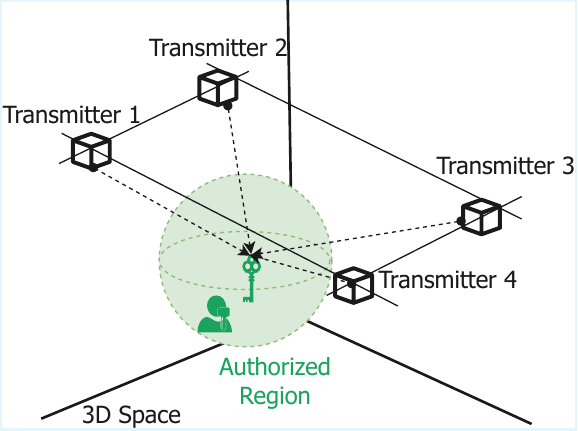}
  \caption{Authorized-Decryption Geometry: Multiple ($\ge 3$) transmitting anchors (or transmitters) define an authorized spatial region in which a receiver observes the intended time-of-flight aligned packet timing pattern and can reconstruct the key for decryption.}
  \label{fig:cryptosystem}
\end{figure}

This intuition motivated our geography inspired defense, \pname: designing a cryptosystem that ties decryption capability to a \emph{geographic region}. Intuitively, we want a mechanism such that, first, a legitimate user located within an authorized region can reconstruct the decryption key automatically, without ever seeing the key in cleartext or manually entering a password, and second, any party outside that region, even if they capture all encrypted traffic, cannot recover the key from the signals they observe.

Rather than transmitting a key as a conventional data payload or shared separately, we exploit precise timing information in ultra-wideband (UWB)~\cite{aiello2003ultra} radio signals. By mapping each byte of a 32-byte SHA-256–derived key to carefully scheduled packet transmission times, we ensure that only receivers experiencing the correct time-of-flight delays (i.e., those physically located in the intended region) reconstruct the intended key. 

We design and implement \pname, a location-dependent cryptosystem, using Ciholas DWETH101~\cite{cuwb_anchors_2020} hardware and design a custom TiCK (Timing-encoded Cryptographic Keying) protocol. The sender encrypts data with AES-256~\cite{nist_fips197_aes} using a password provided by the sender, derives a fixed-length key (32 bytes) by hashing the password using SHA-256~\cite{nist_fips1804_sha}, and encodes that key across a sequence of 33 UWB transmissions (i.e., first transmission marks the beginning and then thirty-two transmission for each byte). The mapping from password encrypted bytes to transmission timestamps is computed using:
(i) a global network time maintained by the server, (ii) a configurable initial offset that aligns with the server’s internal timing windows, (iii) a minimum inter-packet delay to ensure user can process packets in real time, and (iv) per-transmitter time-of-flight offsets corresponding to the authorized decryption region.

A user within the authorized region (as shown on Figure~\ref{fig:cryptosystem}) observes packet arrival times whose time-of-flight align with the pre-defined ``time slots'' for each key byte, allowing it to reconstruct the SHA-256 key and decrypt the AES-encrypted payload. An eavesdropper at a different location, however, will observe shifted time-of-flight patterns; without precise knowledge of both reference timestamps and per-transmitter distances, they derive an incorrect key and fail to decrypt the content.

For \pname's evaluation, we implement a complete working prototype, where we encrypts audio files, transmits the key via UWB packet timings, and automatically decrypts and plays the audio only within the authorized area. The prototype demonstrates that, (i) the decryption password never appears in a data form on the wire, (ii) the authorized region is a configurable \emph{zone} rather than a single point, giving legitimate users spatial tolerance, and (iii) unauthorized user (e.g., eavesdropper) outside this zone do not reconstruct the correct key, even when they capture all packets.

To the best of our knowledge, we present the first exploration of geographically bounded key distribution enabled by wireless transmission, demonstrating that physical location can serve as a first-class primitive rather than merely an external constraint. Instead of treating decryption keys as static secrets that must be guarded and transported, we embed the key into the physical properties of UWB communication, specifically, the time-of-flight patterns observed only within an authorized region. 

\section{Background}

The current encryption standards for industries are AES256 (128 bits or higher), TDES (double-length keys), ECC (160 bits or higher) and ElGamal (1024 bits or higher)~\cite{nist_sp80038a, ansix9_52_2016, pcidss_v4, holz2011surveyssl, koblitz1987ecccipher, elgamal1985publickey, menezes1996handbook}. These encryption standards are secure as the underlying base problem, the discrete log problem (DLP), is intractable and exponentially hard for large primes~\cite{menezes1993discretelog}. Intractable is defined as taking thousands of years to brute force through the function, even for the top five supercomputers of the world.

In industry, the encryption standard used to encrypt data is AES~\cite{nist_fips197_aes}. The encrypted data is then transmitted using any one of the cryptographic network protocols, such as Internet Protocol Security (IPsec)~\cite{rfc4301_ipsec}. However, the key of the encryption is independent of the location and it needs to be transferred to the receiver physically or electronically. Therefore, in this cryptosystem, to make sure the encrypted data can only be decrypted at the authorized location, the key of the encrypted data is associated with the approved location. The system will start the decryption process automatically without the need for the receiver's assistance, thus protecting the password's integrity.

\section{Threat Model.}
We assume an honest sender that encrypts an application payload under an AES-256 key derived from a sender-chosen password via SHA-256, and conveys this key to receivers implicitly through the timing of a 33-packet TiCK UWB transmission burst. An \emph{authorized} receiver is located within a sender-defined approved region, so that its packet time-of-flight matches the pre-compensated propagation terms and it can correctly decode each byte by midpoint rounding over consecutive reception timestamps. The adversary is an \emph{off-region, passive eavesdropper} that can record all over-the-air packets and timestamps, but is physically outside the approved region; therefore its propagation times differ from those assumed by the schedule, shifting inter-arrival gaps and causing slot-decoding errors that yield an incorrect reconstructed SHA stream and, consequently, decryption failure. 

\section{System Design}
\label{sec:system-design}

This section describes \pname's hardware/software stack and the \emph{Timing-encoded Cryptographic Keying (TiCK)} protocol used to convey a SHA-derived key via UWB timing.

\subsection{Hardware}
\label{subsec:hardware}

\pname{} requires a device capable of emitting and relaying ultra-precise timing information. We use \texttt{DWETH101}~\cite{cuwb_anchors_2020} as the transmitter, which provides nanosecond-scale timestamping at the physical layer via its DecaWave UWB chip~\cite{qorvo_dw1000}. We interface the \texttt{DWETH101} with a NetApp server--client system: the server reads the hardware timestamps and exposes them as \emph{Network Time (NT) ticks}. 

\heading{Network Time (NT) ticks.}
An NT tick is an \emph{integer-valued increment of the UWB chip's time counter} (i.e., a discrete clock step), not an operating-system timestamp. Thus, NT ticks form a stable device-local time base derived directly from the physical-layer timestamp counter, and time intervals are represented as tick differences.

\heading{Seconds to NT ticks Conversion.}
Let $\Delta t$ be an elapsed time interval in seconds and $\Delta S$ be its representation in NT ticks. We use these constant,
\begin{equation}
\Delta S = \Delta t \cdot K_f,
\qquad\text{and conversely}\qquad
\Delta t = \frac{\Delta S}{K_f},
\label{eq:sec-ticks-map}
\end{equation}
where $K_f$ is the conversion factor (ticks per second). In our setup,
\begin{equation}
K_f = 975000 \times 65536 = 63{,}897{,}600{,}000
\quad\text{NT ticks/second}.
\label{eq:kf}
\end{equation}
Here $975000$ is a coarse clock rate (Hz) and $65536=2^{16}$ is a fixed-point (Q16) fractional scaling that subdivides each coarse unit into $2^{16}$ finer steps. Therefore,
\[
1\ \text{tick} = \frac{1}{K_f}\ \text{s} \approx 15.65\ \text{ps}.
\]
Interpreting this as a one-way propagation-time resolution yields a nominal distance resolution of $c/K_f \approx 4.69$\,mm per tick.

\heading{From time-of-flight to distance.}
A one-way time-of-flight difference $\Delta t$ corresponds to a path-length difference $\Delta d \approx c\,\Delta t$, where $c$ is the speed of light. Since $1$\,cm corresponds to $\approx 33.3$\,ps of one-way propagation time, tick measurements enable centimeter-scale discrimination. In \pname, however, the \emph{acceptance region} is primarily governed by the protocol slot duration $T_{\text{slot}}$ (details in \autoref{subsubsec:tick}), rather than raw tick quantization alone.

\begin{figure}[t]
  \centering
  \includegraphics[width=\linewidth]{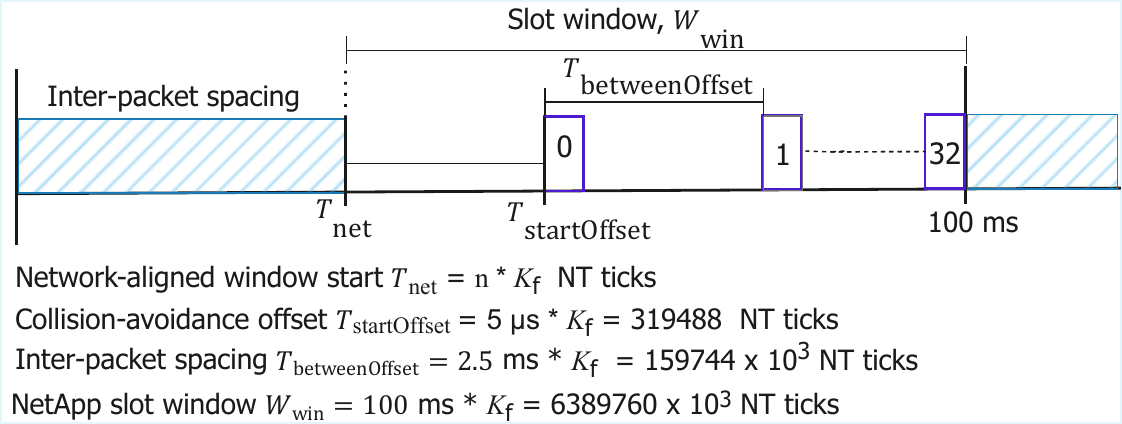}
  \caption{NetApp timing window and TiCK scheduling. TiCK schedules a 33-packet burst (one reference packet and 32 byte-carrying packets) within a single NetApp synchronization window, enforcing the fixed processing gap $T_{\text{betweenOffset}}$ between consecutive packets. The reference packet is transmitted at $T_{\mathrm{tx}}(0)$ with propagation pre-compensation such that, under nominal time-of-flight, it arrives at the approved receiver at the network-aligned epoch $T_{\text{net}}+T_{\text{startOffset}}$, which anchors the subsequent slot-aligned inter-arrival decoding.}
  \label{fig:time-window}
\end{figure}

\heading{\pname: Server.}
The \pname{} server listens to NetApp timing packets and maintains a rolling view of the current time window (illustrated in \autoref{fig:time-window}) so that it can schedule a 33-packet TiCK burst within a single window. Given a sender password, the server computes its SHA-256 hash and encrypts the application payload under an AES key derived from the hash. The server then selects a network-aligned start time $T_{\text{net}}$ and emits the first (reference) TiCK packet at a network-aligned cadence using the TiCK transmit rules.

\heading{\pname: Client (authorized user).}
Authorized receivers continuously listen for TiCK packets, extract reception timestamps (in NT ticks), and reconstruct the SHA-256 bytes via TiCK decoding (refer to \autoref{eq:tick-rx-round}). If no packets arrive for $20$ seconds, the receiver stops listening, finalizes SHA reconstruction, and attempts to decrypt the received encrypted payload. Upon successful decryption, the recovered payload is presented to the user.

\subsection{Software}
\label{subsec:software}

\pname{} comprises: (i) an application-layer payload encryption-decryption cryptosystem (e.g., audio), (ii) AES encryption keyed by a SHA-256-derived value, and (iii) the TiCK protocol that transmits the SHA-256 key bytes via timing. We release the codebase for reproducibility:
\footnote{\url{https://anonymous.4open.science/r/Location_Dependent_Cryptosystem_anon/}}

\heading{Payload pre-processing.}
To demonstrate location-dependent decryption, we use an audio file intended to decrypt \emph{only} at an authorized location selected by the sender. The sender converts \texttt{input.wav} into a frequency-domain representation (principal frequency values) stored for later reconstruction back into audio. Concretely, we use \texttt{ffmpeg} to transform the \texttt{.wav} stream into a frequency-domain representation, which serves as the plaintext payload prior to encryption.

\heading{AES-256 and SHA-256.}
We compute $\text{SHA-256}(\text{password})$ to obtain a 32-byte digest and derive an AES-256 key from this digest. This (i) avoids storing the password in plaintext and (ii) provides a fixed-length key for AES-256 decryption. Encryption uses an initialization vector (IV) and processes the payload in application-layer chunks (AES has a fixed block size of 16 bytes, the chunking above refers to application-level buffering rather than AES block size).

\subsection{Timing-encoded Cryptographic Keying (TiCK) Protocol}
\label{subsubsec:tick}

TiCK transmits the 32 bytes of a SHA-256 output by encoding each byte as an inter-packet timing offset aligned to pre-defined \emph{time slots}.
Because the first decoded byte requires a reference packet, TiCK transmits a total of $33$ packets:
one reference packet (index $0$) followed by $32$ packets that encode the $32$ SHA bytes.

\heading{Timestamps and Units.}
Let $T_{\text{tx}}(n)$ denote the transmit timestamp of packet $n$, and let $T_{\text{rx},i}(n)$ denote the reception timestamp at receiver $i$. Timings are expressed in NT ticks using \autoref{eq:sec-ticks-map}.

\heading{Time Windows and Start Offset.}
NetApp periodically transmits internal timing packets to keep devices synchronized; we call the interval between successive NetApp synchronization emissions a time window. To avoid collisions with higher-priority internal packets, TiCK uses an initial start offset $T_{\text{startOffset}}$.
We use,
\[
T_{\text{startOffset}} = 5~\mu\text{s} = 5\times 10^{-6}\,K_f = 319{,}488\ \text{ticks}.
\]
We define a network-aligned epoch $T_{\text{net}}=nK_f$ for an integer $n$ chosen such that all $33$ transmissions fit inside a single NetApp synchronization window.

\heading{Inter-packet Spacing and Propagation Terms.}
Successive TiCK packets must be separated by a minimum processing gap so that receivers can process a packet and return to listening mode. Let $T_{\text{betweenOffset}}$ denote this fixed gap (empirically $\ge 2.5$ ms in our setup).

Let $T_{\text{distA}}(n)$ denote the expected one-way propagation time (in NT ticks) from the transmitting anchor used for packet $n$ to the \emph{approved location}. This term is used to pre-compensate transmissions so that (for an approved receiver) arrivals are aligned to the intended slot structure.

\heading{Time-slot Mapping.}
Each SHA byte is an integer in $\{0,\dots,255\}$ and is mapped to a \emph{slot number} $S(n)\in\{0,\dots,255\}$. TiCK uses a slot duration $T_{\text{slot}}$ (NT ticks) to convert a slot number into a time offset. The slot duration governs the spatial tolerance of the authorized region: smaller $T_{\text{slot}}$ yields tighter timing tolerance (and potentially smaller spatial acceptance), while larger $T_{\text{slot}}$ yields a larger tolerance.

\begin{figure}[t]
  \centering
  \includegraphics[width=\linewidth]{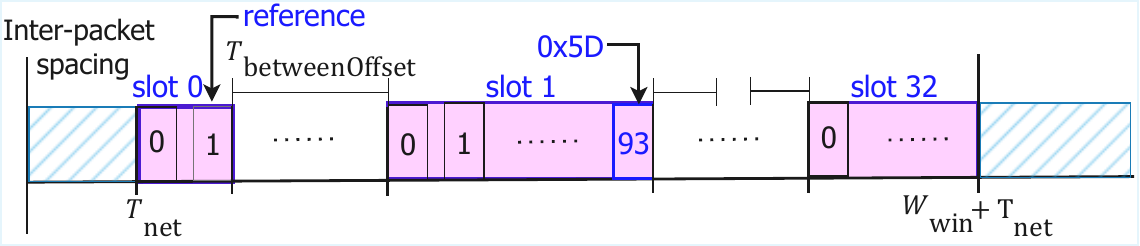}
  \caption{Time-window slotting for timing-encoded bytes. Within a NetApp synchronization window $[T_{\text{net}},\,T_{\text{net}}+W_{\text{win}}]$, TiCK transmits a 33-packet sequence (one reference packet followed by 32 byte-carrying packets). Each SHA-256 byte is mapped to a slot value $S\in\{0,\dots,255\}$ and encoded by scheduling the next transmission after the fixed inter-packet spacing $T_{\text{betweenOffset}}$ plus a slot-dependent offset $(S+\tfrac{1}{2})T_{\text{slot}}$ (illustrated by $S=0x5D=93$), where $T_{\text{slot}}$ controls the timing---and thus spatial---tolerance of the authorized region.
}
  \label{fig:slot-window}
\end{figure}

As a reference, a one-way radius of $r=2$~m corresponds to a propagation time of
$t_{\text{ToF}} = r/c \approx 2/(3\times 10^8)\approx 6.67$~ns, which corresponds to
\[
T_{\text{slot}} \approx t_{\text{ToF}}\cdot K_f \approx 6.67\text{ ns}\cdot 63.9\times 10^9 \approx 426\ \text{ticks}.
\]
This motivates choosing $T_{\text{slot}}$ when constructing the 256-slot timing window (refer to \autoref{fig:slot-window}). This is crucial since a displacement that induces a residual timing shift of roughly $\gtrsim T_{\text{slot}}/2$ can flip a decoded byte.

\heading{Transmit rule: Reference packet ($n=0$).}
TiCK emits a reference packet such that (under nominal propagation to the approved location) it arrives at the approved receiver at $T_{\text{net}}+T_{\text{startOffset}}$:
\begin{equation}
T_{\mathrm{tx}}(0)
=
T_{\mathrm{net}} + T_{\mathrm{startOffset}} - T_{\mathrm{distA}}(0).
\label{eq:tick-tx0}
\end{equation}

\heading{Transmit rule: Byte-carrying packets ($n\ge 1$).}
Packet $n$ (for $n=1,\dots,32$) encodes SHA byte $n-1$ using slot number $S(n-1)\in\{0,\dots,255\}$.
Given the previous transmit time, TiCK schedules:
\begin{equation}
\begin{aligned}
T_{\text{tx}}(n)
&=
T_{\text{tx}}(n-1)
+
T_{\text{distA}}(n-1)
+
T_{\text{betweenOffset}}
\\
&\quad+
\Bigl(S(n-1)+\tfrac{1}{2}\Bigr)\,T_{\text{slot}}
-
T_{\text{distA}}(n).
\end{aligned}
\label{eq:tick-tx}
\end{equation}
The $(\cdot+\tfrac{1}{2})$ midpoint shift places each slot at its center and enables robust rounding at the receiver.

\heading{Receive rule and Decoding.}
Let receiver $i$ experience an one-way propagation time $T_{\text{dist},i}(n)$ for packet $n$.
Then
\[
T_{\text{rx},i}(n) = T_{\text{tx}}(n) + T_{\text{dist},i}(n).
\]
For an approved receiver at the authorized location, $T_{\text{dist},i}(n)=T_{\text{distA}}(n)$ by construction.

Subtracting consecutive arrivals and using \autoref{eq:tick-tx}, an approved receiver obtains
\[
\begin{aligned}
& T_{\text{rx,appr}}(n) - T_{\text{rx,appr}}(n-1) \\
&=
T_{\text{betweenOffset}} + \Bigl(S(n-1)+\tfrac12\Bigr)T_{\text{slot}} .
\end{aligned}
\]
so the receiver can recover $S(n-1)$ via midpoint rounding:
\begin{equation}
\begin{aligned}
\widehat{S}_i(n-1)
&=
\left\lfloor
\frac{
T_{\text{rx},i}(n)-T_{\text{rx},i}(n-1)-T_{\text{betweenOffset}}
}{
T_{\text{slot}}
}
-\tfrac{1}{2}
\right\rceil, \\
&\qquad n=1,\dots,32.
\end{aligned}
\label{eq:tick-rx-round}
\end{equation}
The recovered slot values $\widehat{S}_i(0),\dots,\widehat{S}_i(31)$ are interpreted as the reconstructed SHA-256 bytes.

\heading{Threat model and off-location decoding.}
An eavesdropper outside the approved region generally experiences different propagation times $T_{\text{dist},i}(n)\neq T_{\text{distA}}(n)$, yielding shifted reception timestamps. This produces decoding error in \autoref{eq:tick-rx-round}, causing the reconstructed SHA stream to diverge and decryption to fail.

\begin{figure}[t]
  \centering
  \includegraphics[width=0.8\linewidth]{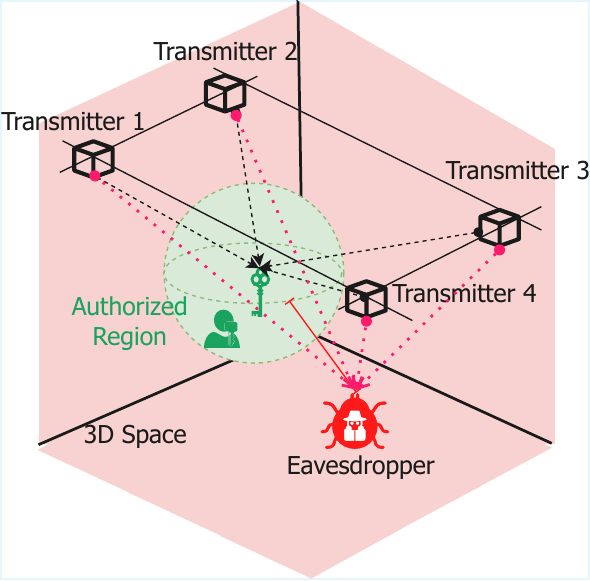}
  \caption{Threat model: authorized receiver vs.\ eavesdropper. An unauthorized receiver outside the approved region observes shifted time-of-flight timing, causing incorrect slot decoding and failure to reconstruct the correct key via \autoref{eq:tick-rx-round}.}
  \label{fig:eavesdropper-system}
\end{figure}

\subsection{\pname{} Working Example: Approved Receiver vs.\ Eavesdropper}
\label{subsec:working-example}

We instantiate TiCK decoding with concrete parameters to illustrate how a small displacement outside the authorized region yields a different decoded SHA byte. 
We refer to \autoref{fig:eavesdropper-system} for an overview of the threat model, illustrating how the four transmitters defines the authorized region and where an eavesdropper may be located.

\paragraph{Parameters.}
We use the NT tick rate $K_f$ from \autoref{eq:kf}, so $1$ tick $\approx 15.65$ ps. We use these values:
\[
T_{\mathrm{net}} = K_f = 63{,}897{,}600{,}000,\quad
T_{\mathrm{startOffset}} = 5~\mu\text{s} = 319{,}488,
\]
\[
T_{\mathrm{betweenOffset}} = 2.5~\text{ms} = 159{,}744{,}000,\quad
T_{\mathrm{slot}}=426.
\]
We consider a single-anchor instance for these two packets and use
\[
T_{\mathrm{distA}}(0)=383,\quad T_{\mathrm{distA}}(1)=667,
\]
and an eavesdropper residual mismatch of
\[
\Delta = 227\ \text{ticks}
\quad(\text{i.e., } T_{\text{dist,eve}}(n)=T_{\text{distA}}(n)+\Delta\ \text{for } n\in\{0,1\}).
\]

\heading{Step 1: Reference packet ($n=0$).}
From \autoref{eq:tick-tx0},
\begin{align}
T_{\mathrm{tx}}(0)
&= T_{\mathrm{net}} + T_{\mathrm{startOffset}} - T_{\mathrm{distA}}(0) \\
&= 63{,}897{,}600{,}000 + 319{,}488 - 383 \notag\\
&= 63{,}897{,}919{,}105, \notag\\\
T_{\mathrm{rx,appr}}(0)
&= T_{\mathrm{tx}}(0) + T_{\mathrm{distA}}(0)
= 63{,}897{,}919{,}488,\\
T_{\mathrm{rx,eve}}(0)
&= T_{\mathrm{tx}}(0) + (T_{\mathrm{distA}}(0)+\Delta)
= 63{,}897{,}919{,}715.
\end{align}

\heading{Step 2: First byte packet ($n=1$).}
Let the first SHA byte be $S(0)=109$ (hex $0x6D$). Using \autoref{eq:tick-tx} with $(S(0)+\tfrac12)T_{\text{slot}}$, we get $109.5\times 426=46{,}647$,
\begin{align}
T_{\mathrm{tx}}(1)
&=
T_{\text{tx}}(0)
+ T_{\text{distA}}(0) + T_{\text{betweenOffset}} \notag\\\
&+ 46{,}647
- T_{\text{distA}}(1)
\\
&=
63{,}897{,}919{,}105
+ 383
+ 159{,}744{,}000 \notag\\
&+ 46{,}647- 667
= 64{,}057{,}709{,}468, \notag\\
T_{\mathrm{rx,appr}}(1)
&= T_{\mathrm{tx}}(1) + T_{\mathrm{distA}}(1)
= 64{,}057{,}710{,}135,\\
T_{\mathrm{rx,eve}}(1)
&= T_{\mathrm{tx}}(1) + (T_{\mathrm{distA}}(1)+\Delta)
= 64{,}057{,}710{,}362.
\end{align}

\heading{Decoding.}
Apply \autoref{eq:tick-rx-round}:
\begin{align}
\widehat{S}_{\mathrm{appr}}(0)
&=
\left\lfloor
\frac{T_{\text{rx,appr}}(1)-T_{\text{rx,appr}}(0)-T_{\text{betweenOffset}}}{T_{\text{slot}}}
-\tfrac12
\right\rceil \\
&=
\left\lfloor
\frac{46{,}647}{426}-\tfrac12
\right\rceil
=109\ (=0x6D),\notag\\
\widehat{S}_{\mathrm{eve}}(0)
&=
\left\lfloor
\frac{T_{\text{rx,eve}}(1)-T_{\text{rx,eve}}(0)-T_{\text{betweenOffset}}}{T_{\text{slot}}}
-\tfrac12
\right\rceil
\\
&=
\left\lfloor
\frac{46{,}647}{426}-\tfrac12 + \frac{\Delta}{426}
\right\rceil. \notag
\end{align}
Since $\Delta=227$ ticks and $T_{\text{slot}}/2=213$ ticks, the eavesdropper’s residual exceeds the midpoint threshold, yielding
\[
\widehat{S}_{\mathrm{eve}}(0)=110\ (=0x6E)\neq 109.
\]
In physical units, $\Delta t=\Delta/K_f\approx 227/63{,}897{,}600{,}000\approx 3.55$ ns, corresponding to a one-way path-length difference of $\Delta d\approx c\Delta t\approx 1.06$ m. This residual is sufficient to push the received timestamp across a slot boundary under midpoint decoding, flipping the decoded byte outside the authorized region.
\section{Results}

\heading{Cryptosystem Simulator.}
\label{subsubsec:simulator}
We developed a proof-of-concept simulator to validate that the transmit/receive rules (\autoref{eq:tick-tx0}, \autoref{eq:tick-tx}, \autoref{eq:tick-rx-round}) reliably convey the intended SHA bytes under controlled conditions. The simulator uses TCP as a stand-in transport and introduces a noise term $T_{\text{noise}}$ (in NT ticks) to model measurement effects.

\heading{Noisy decoding rule.}
We modify \autoref{eq:tick-rx-round} by adding $T_{\text{noise}}$ to the inter-arrival difference:
\begin{equation}
\begin{aligned}
\widehat{S}(n-1)
&=
\left\lfloor
\frac{
\bigl(T_{\text{rx}}(n) - T_{\text{rx}}(n-1)\bigr)
- T_{\text{betweenOffset}}
+ T_{\text{noise}}
}{
T_{\text{slot}}
}
-\tfrac{1}{2}
\right\rceil, \\
&\qquad n=1,\dots,32.
\end{aligned}
\label{eq:tick-rx-noise}
\end{equation}

\heading{Simulator workflow.}
The simulator server takes an audio file, a password, and anchor-to-approved-location distances (converted to ToF in ticks). The server computes SHA-256 on the password, encrypts the audio payload (refer to \autoref{subsec:software}), divides the ciphertext into application-layer chunks, and sends each chunk along with the simulated transmit timestamp. The first network-aligned epoch is set to $T_{\text{net}}=nK_f$ NT ticks (refer to \autoref{eq:kf}); the TiCK cadence begins with the reference packet using \autoref{eq:tick-tx0} and continues with \autoref{eq:tick-tx}.

The simulator client receives packets, uses the received timestamps to recover the 32-byte SHA value via \autoref{eq:tick-rx-noise}, and then attempts to decrypt the ciphertext. If the recovered SHA matches the sender-side SHA, the client reconstructs and plays the intended audio; otherwise, decryption yields noise, indicating failure. These experiments validate that TiCK remains deployable under controlled noise, supporting feasibility over a UWB timing medium.

The \pname demonstrated all requirements as stated during the introduction discussion. \pname was able to transfer the SHA of the password by encoding it into time-of-flight reception timing pattern. The approved location was a space rather than a point. This ensured that the user had a certain degree of freedom and the location had a certain tolerance.

\section{Discussion}

This study demonstrates that physical location can be elevated from an external policy constraint to a first-class input to key reconstruction: \pname encodes a 32-byte SHA-256 digest into UWB packet timing such that correct decoding (and hence decryption) is only achievable within an authorized spatial region. Concretely, the transmitter maps each SHA byte to a slot value $S(k)\in\{0,\dots,255\}$ and schedules packet inter-arrival gaps using the TiCK slot duration $T_{\text{slot}}$ and fixed processing gap $T_{\text{betweenOffset}}$; a receiver reconstructs the digest by decoding slot values from consecutive reception timestamps $T_{\text{rx},i}(\cdot)$ via midpoint rounding (cf.\ \autoref{eq:tick-rx-round}). While the prototype validates feasibility, several deployment and security questions naturally arise.

\heading{Location--utility trade-off via $T_{\text{slot}}$.}
A central design knob is the slot duration $T_{\text{slot}}$, which governs the temporal separation between adjacent slot centers and therefore the spatial tolerance of the authorized region. Under midpoint decoding, a receiver’s decoded byte changes when the residual timing shift induced by off-location propagation and measurement noise exceeds approximately $T_{\text{slot}}/2$ (in NT ticks). Thus, increasing $T_{\text{slot}}$ increases the tolerated timing (and spatial) variation for legitimate receivers, reducing false rejections under jitter and multipath, but simultaneously reduces the effective ``margin'' separating adjacent slot decisions, making off-region reconstruction easier. In addition, TiCK imposes real-time scheduling constraints---most notably the minimum inter-packet processing gap $T_{\text{betweenOffset}}$---which limits throughput (33 packets per key epoch) and can affect robustness under load.

\heading{\textit{Does the security argument extend beyond a passive off-region eavesdropper?}}
The current rationale focuses on a passive adversary who records all packets but experiences different time-of-flight terms (i.e., $T_{\text{dist},\text{eve}}(n)\neq T_{\text{distA}}(n)$), causing shifted reception timestamps and incorrect slot decoding under \autoref{eq:tick-rx-round}. For stronger real-world assurance, evaluation should explicitly consider adversaries that can \emph{actively manipulate} what the receiver observes, including:
(i) \emph{replay} (reproducing a previously observed TiCK timing schedule within a later window),
(ii) \emph{relay/wormhole} (forwarding or repeating signals in real time to emulate in-region reception),
(iii) \emph{anchor compromise} (a malicious transmitter emitting attacker-chosen timing sequences), and
(iv) \emph{collusion} (multiple off-region receivers combining observations to reduce uncertainty about slot boundaries).

Practical mitigations include time-bounded sessions that bind decoding to a short validity window anchored at $T_{\text{net}}$ (rejecting schedules outside the intended window), per-session randomness $\eta$ folded into key derivation (e.g., $K_{\text{AES}}$) so that old captures do not decrypt new payloads, and transmitter authentication of the intended schedule (e.g., cryptographic tags over ($T_{\text{net}}$, $T_{\text{startOffset}}$, $T_{\text{slot}}$, $T_{\text{betweenOffset}}$, $\{S(k)\}$)) so receivers can reject attacker-crafted timing. These steps follow standard security methodology: explicitly enumerate attacker capabilities and stress-test the system under those capabilities rather than relying on a passive threat model.

\heading{\textit{How should $T_{\text{slot}}$ and the authorized region be selected in a principled way?}}
The design links $T_{\text{slot}}$ to region size through one-way timing tolerance (meters divided by $c$) and uses 256 slots corresponding to byte values. A principled selection should incorporate (i) measurement noise and clock jitter in the observed inter-arrival gaps $\Delta T_i(n)\triangleq T_{\text{rx},i}(n)-T_{\text{rx},i}(n-1)$, and (ii) application requirements such as an acceptable false-reject rate for in-region users. One robust approach is to view slot selection as an optimization problem: choose $T_{\text{slot}}$ to minimize off-region key recovery success subject to a constraint on in-region decoding success under the empirical noise distribution. In this sense, $T_{\text{slot}}$ acts like a ``physical tolerance budget'' that must be justified experimentally, analogous in spirit to how security systems tune a privacy/utility knob based on measured trade-offs.

\heading{\textit{What are the scalability limits of a timing-encoded key slots?}}
TiCK transmits 32 SHA bytes using a 33-packet schedule (one reference packet plus 32 byte-carrying packets), so the per-session key distribution overhead is fixed and non-trivial when keys are refreshed frequently or many receivers are served. Two natural scalability directions are: (i) amortize the in-region setup by using the reconstructed digest as a short-lived seed from which multiple content keys are derived via a key hierarchy, and (ii) introduce redundancy selectively (e.g., error-correcting codes across bytes or repeated transmission of high-error bytes) only when channel conditions degrade. These improvements follow common robustness principles: amortize expensive secure establishment, and add redundancy where empirical failure modes concentrate.

\section{Future Work}

Even with these security features, this system can be defeated by implementing many secondary anchors and using them to triangulate the approved location. The secondary anchors are used to determine which anchor sent which packets and their relative times of flight. Moreover, an important feature can be developed called a rolling key encryption system. With this, the receiver must immediately transmit a specific packet upon receiving an encrypted packet. The server will determine whether the timestamp for this specific packet is from the approved location. If not, the server will change the encryption key and re-transmit. If so, the next packet will be sent. This will continue until all of the packets have been transmitted and the cryptosystem can function normally.

\section{Related Work}
\label{sec:related-work}

\heading{Geography and Location-bound Access Control.}
Early work on geo-encryption proposed binding decryption to a geographic region by incorporating position into key construction, so that ciphertext becomes unusable outside an authorized area. Scott et al. formalize this concept and discuss applications such as controlled content distribution and location-authenticated workflows~\cite{scott2003geoencryption_gpsworld,scott2003location_based_encryption}.

\heading{Position as a first-class Cryptographic Factor.}
In contrast to geo-encryption, which often assumes an external positioning oracle, and standard secure-ranging primitives (which primarily aim to measure distance securely), a location-dependent cryptosystem elevates physical location into the key distribution mechanism itself. This viewpoint aligns with the broader goal of making position an explicit security principal, while inheriting the core challenge emphasized by secure ranging and distance-bounding research: the system must remain robust against relay/distance manipulation at the physical layer~\cite{ieee802154z2020,leu2022ghostpeak,anliker2023clocks,wang2025provcreator,Mukherjee2025ZREx}.

\heading{Secure Ranging with UWB and Standardization.}
Modern location-aware systems increasingly rely on UWB time-of-flight ranging; IEEE 802.15.4z enhances UWB PHY/MAC specifically to improve ranging integrity (e.g., secure timestamp sequences)~\cite{ieee802154z2020}. However, recent work shows that implementation details and physical-layer effects can still enable practical distance manipulation. In particular, Ghost Peak demonstrates practical distance-reduction attacks against HRP UWB ranging systems in realistic settings~\cite{leu2022ghostpeak}, and subsequent analysis highlights how clock imperfections can fundamentally weaken secure ranging assumptions~\cite{anliker2023clocks}.
\section{Conclusion}

\pname{} demonstrates that physical location can be elevated from an external policy constraint to a first-class input to key reconstruction by decoding a timing-encoded, SHA-derived decryption key from carefully scheduled UWB packets. Receivers within the authorized region reconstruct the correct key and transparently decrypt the audio payload, while off-region listeners, even if they record every packet, observe time-of-flight shifts that push arrivals across slot boundaries and yield an incorrect key. Because the system transmits only encrypted payload data and timing structure (never the password itself), it reduces key-exposure risk; and because decoding and decryption occur automatically upon reception, it also reduces operational friction and key-handling errors.

More broadly, this prototype shows that cryptographic access control can be strengthened by coupling it to the physical layer: radio geometry, precise timing, and spatial constraints provide an additional axis of enforcement beyond purely digital policies. Future work should (i) quantify the reliability--security trade-off induced by the slot duration $T_{\text{slot}}$ under real-world noise and multipath, (ii) extend evaluation to stronger adversaries (e.g., replay, relay/wormhole, and compromised transmitters), and (iii) explore integrations with authenticated transmitters, hardware-backed key storage, and multi-factor attestation. Overall, \pname{} offers both a concrete design pattern and an initial evidence base for geographically bounded decryption as a practical complement to traditional cryptographic protections.

\subsection{Impact Statement}
\label{subsec:impact}

\pname is a research prototype aimed at enabling secure, location-dependent access control for encrypted content. The system does not introduce direct social or environmental risks beyond those typical for short-range wireless systems. Additionally, the underlying hardware is FCC-approved, indicating compliance with applicable RF exposure and emissions requirements.

\section{Acknowledgments}

We gratefully acknowledge the support and sponsorship of Mr. Mike Ciholas and the guidance of our Design Advisor, Mr. Justin Bennett. We also thank our Academic Advisor, Dr. Donald Roberts, and Software Advisor, Mr. Tim DeBaillie, for their valuable technical feedback and support throughout the system design and implementation.

\bibliographystyle{ijcaArticle}
\bibliography{BibTex-Sample}

\end{document}